# Title: Spectrally resolved helium absorption from the extended atmosphere of a warm Neptune-mass exoplanet

**Authors:** R. Allart[1]*, V. Bourrier[1], C. Lovis[1], D. Ehrenreich[1], J.J. Spake[2], A. Wyttenbach[1,3], L. Pino[1,4,5], F. Pepe[1], D.K. Sing[2,6], A. Lecavelier des Etangs[7]

**Affiliations:**

[1]Observatoire Astronomique de l'Université de Genève, Université de Genève, Chemin des Maillettes 51, 1290 Versoix, Switzerland.

[2]Astrophysics Group, School of Physics, University of Exeter, Stocker Road, Exeter, EX4 4QL, UK.

[3]Leiden Observatory, Leiden University, Postbus 9513, 2300 RA Leiden, The Netherlands

[4]Dipartimento di Fisica e Astronomia 'Galileo Galilei', Univ. di Padova, Vicolo dell'Osservatorio 3, Padova I-35122, Italy

[5] Anton Pannekoek Institute for Astronomy, University of Amsterdam, Science Park 904, 1098 XH Amsterdam, The Netherlands

[6]Department of Earth and Planetary Sciences, Johns Hopkins University, Baltimore, MD, USA.

[7]Institut d'Astrophysique de Paris, CNRS, UMR 7095, Sorbonne Université, 98 bis boulevard Arago, F-75014, Paris, France

*romain.allart@unige.ch

**Abstract:** Stellar heating causes atmospheres of close-in exoplanets to expand and escape. These extended atmospheres are difficult to observe because their main spectral signature – neutral hydrogen at ultraviolet wavelengths – is strongly absorbed by interstellar medium. We report the detection of the near-infrared triplet of neutral helium in the transiting warm Neptune-mass exoplanet HAT-P-11b using ground-based, high-resolution observations. The helium feature is repeatable over two independent transits, with an average absorption depth of 1.08±0.05%. Interpreting absorption spectra with 3D simulations of the planet's upper atmosphere suggests it extends beyond 5 planetary radii, with a large scale height and a helium mass loss rate $\lesssim 3 \times 10^5$ g·s$^{-1}$. A net blue-shift of the absorption might be explained by high-altitude winds flowing at 3 km·s$^{-1}$ from day to night-side.

**One Sentence Summary:** The resolved absorption signature of helium from the warm Neptune HAT-P-11b traces the extension of its upper atmosphere and allow us to constrain its properties.

**Main Text:**

HAT-P-11b is a transiting, warm Neptune-class exoplanet (27.74 ± 3.11 Earth masses, 4.36 ± 0.06 Earth radii) orbiting its star in 4.89 days (*1–3*). Its orbit is near the edge of the evaporation desert, a region at close orbital distances characterized by a lack of observed Neptune-mass exoplanets (*4, 5*). The evaporation desert can be explained as the result of heating planetary atmospheres by stellar radiative flux: planets which are insufficiently massive lose their gaseous atmospheres through its expansion and hydrodynamic escape (*6, 7*). The upper atmosphere of planets in mild conditions of irradiation, like HAT-P-11b, could extend without being subjected to substantial loss and yield deep transit. The low density of HAT-P-11b and the detection of water in its atmosphere (*8*) suggest a hydrogen-helium rich atmosphere clear of aerosols down to an altitude corresponding to 1 millibar atmospheric pressure.

We observed 2 transits of HAT-P-11b with the CARMENES (Calar Alto high-Resolution search for M dwarfs with Exoearths with Near-infrared and optical Échelle Spectrographs, (*9*)) instrument on the Calar Alto 3.5 meter telescope on 2017 August 7 (Visit 1) and 12 (Visit 2). CARMENES consists of two high-resolution spectrographs covering parts of the visible (5,200−9,600 Å) and near-infrared (9,600−17,100 Å) domains with spectral resolving powers of ~95,000 and ~80,000, respectively. We analyse data from the near-infrared channel. The data are automatically reduced with the CARMENES Reduction and Calibration pipeline (*10*), which applies a bias, flat-field and cosmic ray correction to the raw spectra. A flat-relative optimal extraction (FOX; (*11*)) and wavelength calibration (defined in vacuum) are then applied to the spectra (*12*). HAT-P-11 was observed for 6 and 5.8 hours in visits 1 and 2, respectively, in 53 and 51 exposures, each of 408 s. Among these spectra, 19 and 18 were obtained during the 2.4-hour duration of the planetary transit in visits 1 and 2, respectively (*13*).

During a transit, the atmosphere of a planet blocks a fraction of the stellar light at a given wavelength, depending on its structure and composition. We retrieved the near-infrared transmission spectrum of the exoplanet atmosphere by calculating the ratio between the in-transit spectra and an out-of-transit master spectrum (Fig. S1), representing the unocculted star. The out-of-transit master for each visit was determined by co-adding spectra taken before and after transit (*13*). Because of the change in radial velocity arising from the planet's motion, the spectrum of its atmosphere experiences a spectral shift during the transit. Transmission spectra are calculated for each in-transit exposure, offset in wavelength to the planet's rest frame, and co-added (*14–16*) to search for absorption from the planet atmosphere. HAT-P-11b has an eccentric orbit (eccentricity $e = 0.26$) causing the planetary radial velocity to increase from $-36$ km·s$^{-1}$ to $-24$ km·s$^{-1}$ during the transit. As a result, any absorption signatures from the planet atmosphere are expected to be blue-shifted with respect to their rest wavelengths in the stellar reference frame. This helps to distinguish between signals with stellar or planetary origins. The signal-to-noise ratio also increases, as the planet absorption is offset from the equivalent stellar line, unlike planets on circular orbits ((*13*) and Fig. S2). A search for atmospheric absorption features in excess of the planetary continuum absorption signal, optical transit depth ~ 3,400 ppm, revealed absorption in the near-infrared He I triplet (10,832.06, 10,833.22 and 10,833.31 Å in vacuum), displayed in Fig. 1 and Fig. 2. The He I triplet originates from a transition between the $2^3$P state and the metastable $2^3$S state, which can be populated by recombination from the singly-ionized state or by collisional excitation from the ground state (*17*). The triplet is spectrally and temporally resolved during the transit due to the high spectral resolution and fast cadence of the observations. The two strongest lines of the triplet are blended, whereas the third, weakest (and bluest) line is resolved from the two others. These transitions occur in a spectral region devoid of strong water absorption lines or OH emission lines from Earth's atmosphere (*13*). The spectral region is also devoid of strong stellar absorption features ((*13*) and Fig. S1).

Absolute fluxes cannot be determined from ground-based high-resolution spectra because of the variability of Earth's atmosphere and light losses at the spectrograph entrance. Although this does not prevent the detection of spectrally localized absorption features arising from the planet atmosphere, it does mean these observations are not sensitive to any continuum occultation by the atmosphere. To highlight the excess atmospheric He I absorption from HAT-P-11b and to allow comparison with simulations (see below), we rescaled the continuum of each individual transmission spectrum using the theoretical transit white light curve of the planet (Table S1, (*13*)). Fig. 2 shows the rescaled transit light curve integrated over the spectral range 10,832.84−10,833.59 Å, including the most significant excess atmospheric absorption of 1.08±0.05% (21σ, where σ is the standard deviation). The absorption signature is centered on the He I triplet in the planet reference frame, occurs during the planet transit, and is repeatable over two visits (0.82±0.09% in visit 1 and 1.21±0.06% in visit 2), so it arises from helium in the atmosphere of HAT-P-11b (Fig. 1 and Fig. S2). The difference in absorption between the two visits could arise from variability in the size of the atmosphere or its helium density. The peak of the resolved helium absorption profile reaches ~1.2%, corresponding to an equivalent opaque radius of ~2.29 planetary radii.

We interpret the observations of HAT-P-11b using 3D simulations of its atmosphere using the EVaporating Exoplanets (EVE) code ((*18*, *19*); see (*13*) for a detailed description). EVE was used to generate theoretical spectra after absorption by the planet and its upper atmosphere, accounting for limb-darkening, for the partial occultation of the stellar disk during ingress/egress and for 3D effects linked to the atmospheric structure. The upper atmosphere consists in the thermosphere, the layer heated by the stellar irradiation, and the exosphere, the above layer in which the gas becomes collisionless (*13*). The thermosphere is parameterized with an isotropic, hydrostatic density profile defined by the ratio between the temperature of the thermosphere and its mean atomic weight, $T_{th}/\mu$. We included a constant upward velocity $v_{th}$ to account for the bulk expansion of the thermosphere driven by the stellar extreme ultraviolet (XUV) irradiation. This expansion can lead to substantial mass loss, so we modeled the exosphere of HAT-P-11b by releasing metastable helium atoms at the top of the thermosphere (*13*). The altitude $R_{trans}$ of the thermopause (the thermosphere/exosphere boundary, also known as the exobase) is a free parameter in the fitting, as is the mass loss rate of metastable helium $\dot{M}_{\text{He}^*}$. Monte-Carlo particle simulations are used to compute the dynamics of the atmosphere under the influence of the planet and star gravity, and the stellar radiation pressure. The density profile of metastable helium in the thermosphere is scaled so that it matches the density of exospheric metaparticles at the thermopause. We assume that the low densities in the collision less exosphere prevent the formation of additional metastable helium atoms, or their de-excitation through collisions (*17*).

The theoretical spectra are oversampled in time and wavelength compared to the CARMENES observations. We therefore convolve the output with the instrumental response, resampled over the CARMENES wavelength scale, and averaged within the time windows of each observed exposure. The theoretical and observed time-series spectra were compared over visits 1 and 2 together (103 exposures), limiting the fitting to the spectral range 10,826−10,834 Å (139 pixels, defined in the star rest frame) to avoid contamination by Earth atmosphere. We calculated a grid of simulations as a function of the four free parameters in the model, using $\chi^2$ as the merit function (Fig. S4).

The exploration of the model parameter space reveals a broad $\chi^2$ minimum ($\chi^2 \sim 6,130$ for 14,178 datapoints). The best-fitting thermopause altitudes extend between 5 $R_p$ and the Roche Lobe (the limit of the gravitational influence of the planet, at 6.5 $R_p$). The $T_{\text{th}}/\mu$ values indicate high temperatures and/or low mean atomic weight ($T_{\text{th}}/\mu \geq 24,000$ K·amu$^{-1}$), suggesting a large fraction of ionized gas and free electrons. The width of the absorption signature is dominated by thermal broadening, but the upward expansion of the thermosphere could play a role with $v_{\text{th}}$ up to 10 km·s$^{-1}$, in the range of values predicted for HAT-P-11b (*20*). Comparison between our best-fitting signature from a radially expanding thermosphere and the observed absorption profile reveals that it is symmetrical but blue-shifted (Fig 3). Including zonal winds flowing from day- to night-side in our best-fitting models provides a better match to the observed absorption profile ($\chi^2$ = 6,121) for velocities of ~3 km·s$^{-1}$ (Fig. S4). 2D hydrodynamical simulations (*21*) have shown that such winds can form at high altitudes in the extended atmospheres of giant planets.

Our results suggest a negligible contribution from the exosphere, with $\dot{M}_{He^*}$ below $3 \times 10^5$ g·s$^{-1}$. This is consistent with the spectral symmetry of the observed absorption profile near the He I triplet, and the symmetry of the time series absorption around the transit center (Fig. 2). These properties demonstrate that absorption from HAT-P-11b arises mostly from spherical layers of gas likely to be still gravitationally bound to the planet. The absence of post-transit absorption, or a strong absorption signal blueward of the helium transitions, rule out an extended tail of helium trailing the planet. This is unlike the elongated hydrogen exosphere detected around GJ 436 b (*22*, *23*), a warm Neptune with similar density to HAT-P-11b. Our best-fitting models yield densities of metastable helium ~ 10 cm$^{-3}$ at altitudes between 5 and 6.5 $R_p$, within the range of values simulated for GJ 436 b at similar altitudes (*17*).

Our best-fitting simulation in Fig. 4 and S5 illustrates how the radiative environment from the host star (spectral type K4) influences the exosphere of HAT-P-11b. Helium atoms in the shadow of the planet keep move on their original escape trajectory (determined by the orbital velocity of the planet when they escaped), until they radiatively de-excite with a lifetime of 131 min (*24*). Outside of the planet shadow, helium atoms are blown away faster than this lifetime by the strong stellar radiation pressure. It is much stronger at the helium triplet wavelength (~10,833 Å) than for the hydrogen Lyman-α wavelength (1,215.7 Å)), because of the brighter near-infrared continuum. Radiation pressure on metastable helium atoms escaping HAT-P-11b is higher than the gravitational pull of the star by a factor of ~90, whereas it reaches a maximum of ~5 for the hydrogen exospheres of planets around G- and K-type stars (e.g.,(*18*)). However, the low photoionization threshold of metastable helium atoms implies that their lifetime is only 2.4 min at the orbital distance of HAT-P-11b, explaining why we do not observe an extended comet-like tail trailing the planet. There are therefore extended upper atmospheres around both warm Neptunes HAT-P-11b and GJ 436 b. Although they have similar mass and radius, the different spectral types and XUV emission of their host stars (K- and M-type, respectively) are expected to produce different structures for their upper atmospheres. The presence of helium at high altitudes around HAT-P-11b nonetheless suggests that large amounts of hydrogen could be escaping into its exosphere.

Theoretical models (*17*, *25*) have predicted the metastable He I triplet can be used to trace atmospheric evaporation. Because the 10,833 Å He I triplet is not absorbed by the interstellar medium, it allows probing planetary systems farther from Earth than the H I Lyman-α at 1,215 Å (*26*). While early searches were unsuccessful due to instrumental limitations (*27*), an unresolved detection of metastable helium on the inflated gas giant WASP-107b has been achieved with the *Hubble Space Telescope (HST)* (*28*). Because of the low spectral resolution of *HST*, the helium triplet in WASP-107b was covered with just one pixel. We have calculated that observations of HAT-P-11b helium atmosphere with the *James Webb Space Telescope (JWST)* would measure the triplet with a high sensitivity but over just two pixels (*13*). High resolution spectrographs have the ability to spectrally resolve the He I transitions, aiding in the separation of planetary from stellar signals. As shown by our results (Fig. 2) and (*29*)**,** resolved observations of the He I triplet provide additional constraints on the extended atmospheres of exoplanets, from their thermosphere to exosphere.

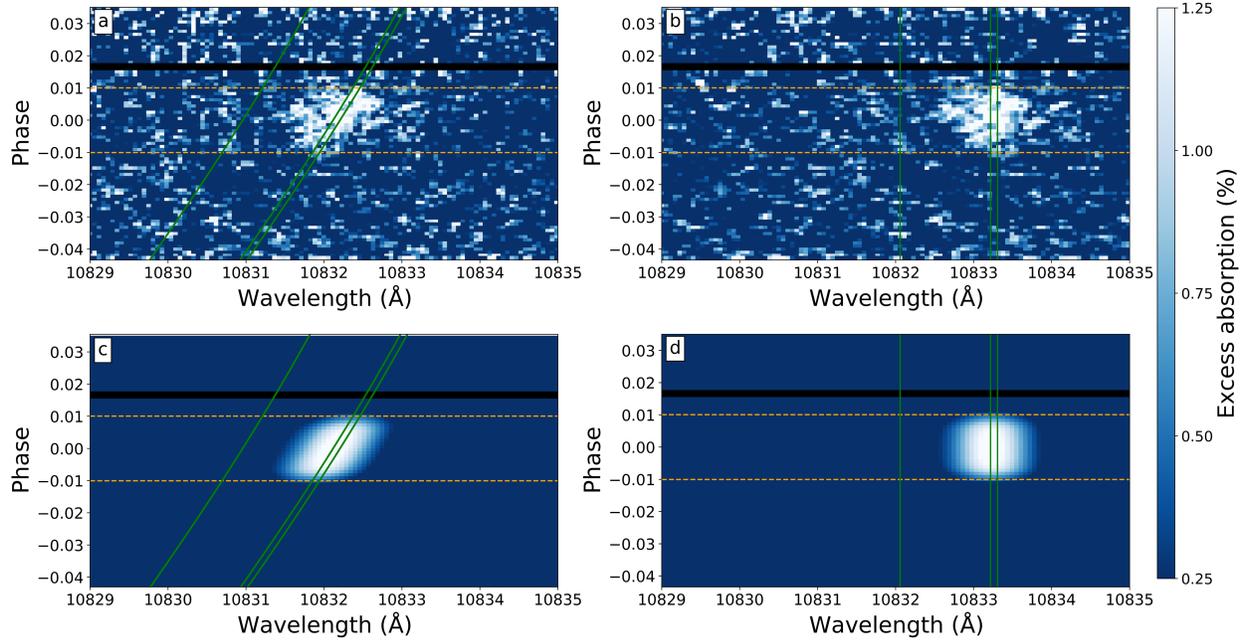

**Fig. 1. He I transmission spectra of HAT-P-11b as a function of orbital phase,** plotted in the star rest frame (*panel a*) and in the planet rest frame (*panel b*). Horizontal orange dotted lines correspond to the beginning and end of transit. Green lines show the three He I transitions in the planet rest frame. Excess atmospheric absorption is visible as a white signal centered on the He I transitions, following the planetary motion, and occurring only during the transit. *Panels c and d* show the equivalent simulated maps for our best-fitting atmospheric model (*13*).

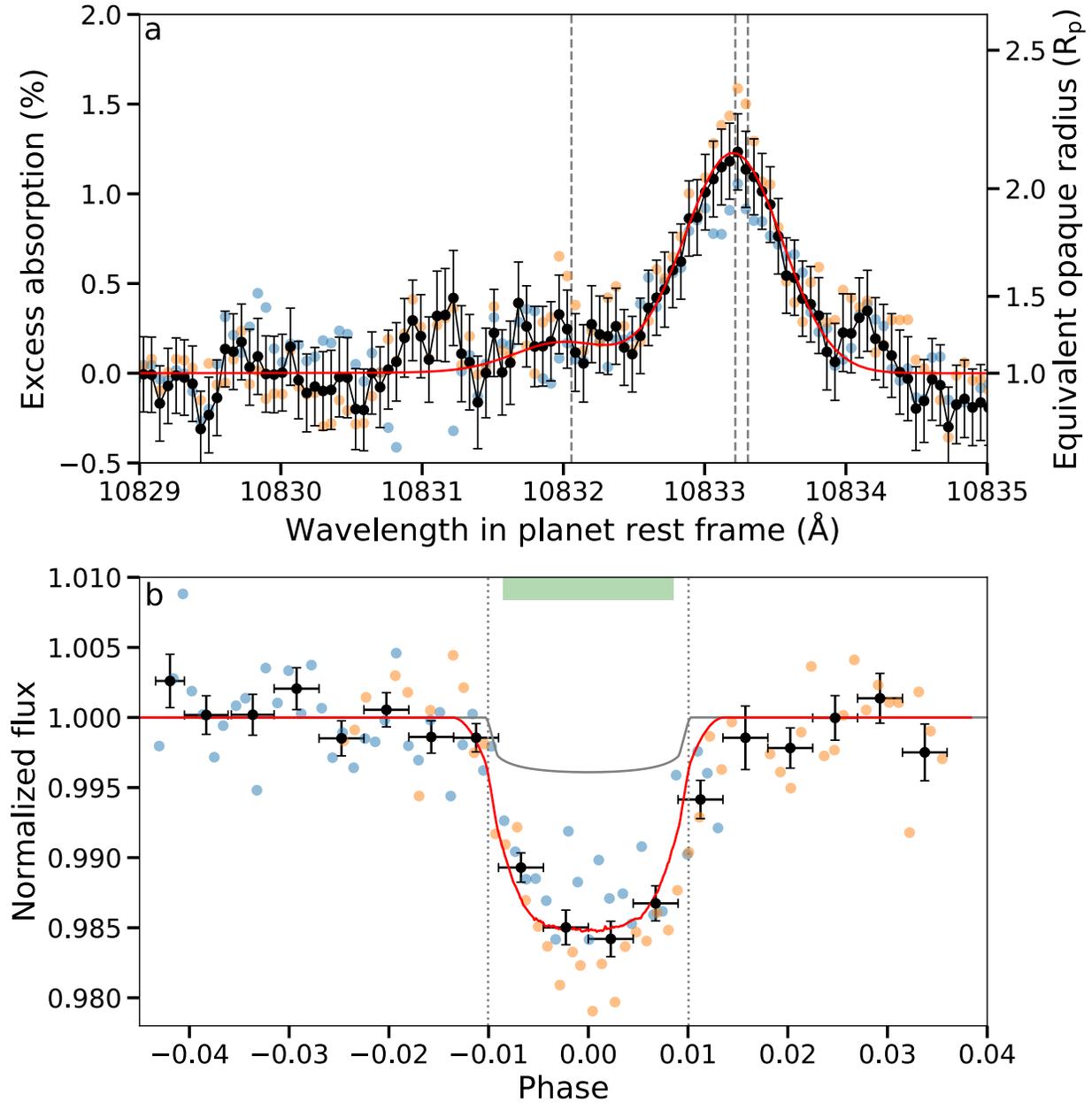

**Fig. 2. Average He I transmission spectrum of HAT-P-11b in the planetary rest frame and transit light curve.** *Panel a:* Transmission spectra in Visit 1 (blue) and Visit 2 (orange), showing the absorption signature centered on the He I triplet (rest wavelengths shown as dashed black lines). The black points show the weighted average over the two visits, and the red line is our best-fitting model. Wavelengths are in the planet rest frame. *Panel b:* Light curves derived from the spectra in the top panel normalized to the expected planetary continuum absorption and integrated over 10,832.84–10,833.59 Å. Plotting symbols are the same as in (a), whilst the theoretical planet light curve without helium absorption is shown in grey. The black light curve was binned in phase. The green band is the time window (−1 h to +1 h) used to produce the average spectrum in (a). Vertical grey dashed lines correspond to the beginning and end of the transit.

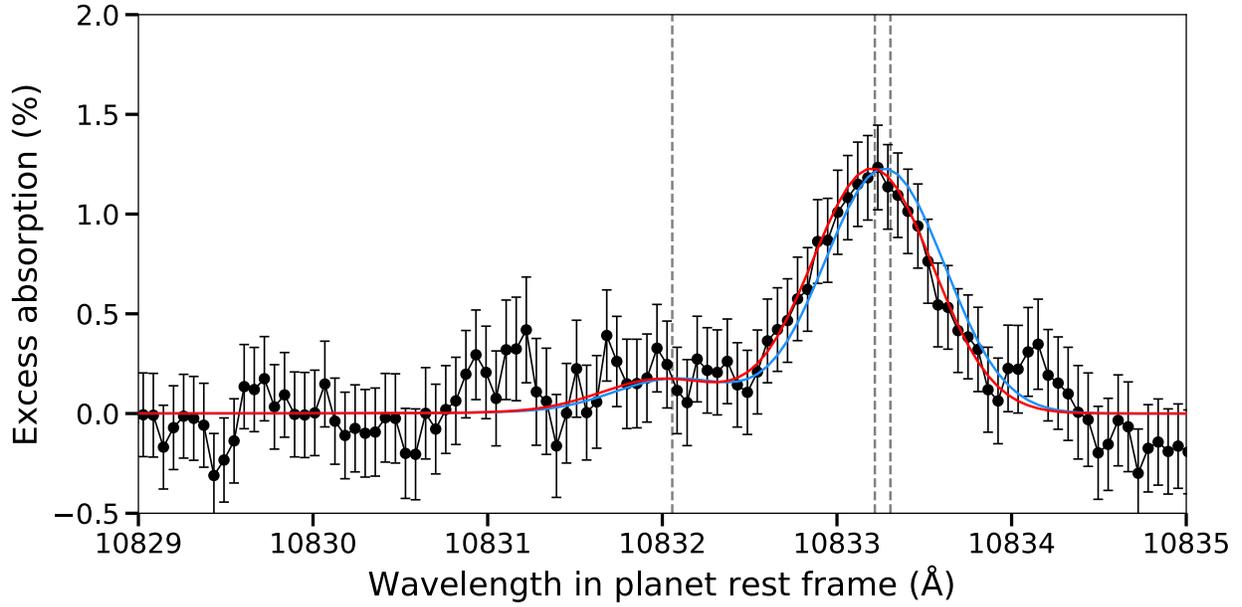

**Fig. 3. Contribution of zonal winds to the He I absorption signature from HAT-P-11b.** The black points are the observed average over the two nights, as shown in Fig. 2a. The blue curve corresponds to a radially expanding thermosphere, while the red curve (shown in Fig. 2a) is blue-shifted by the additional contribution of zonal winds flowing from day- to night-side at 3 km·s$^{-1}$.

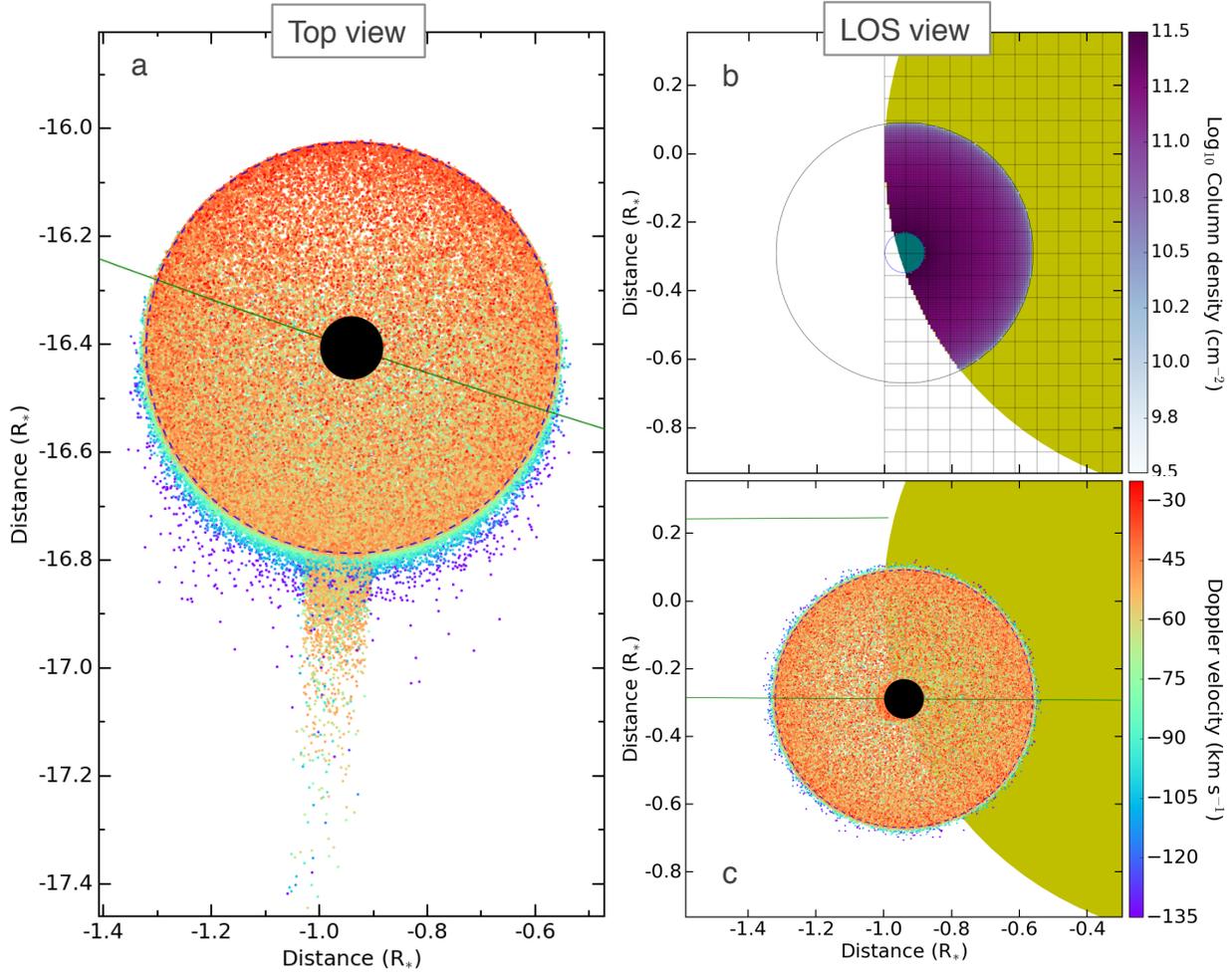

**Fig. 4. The best-fitting EVE simulation of the HAT-P-11b helium absorption time-series.** The system is shown during transit ingress; egress and mid-transit are shown in Fig. S5. Distances are defined with respect to the star center. *Panel a*: View of the exosphere from above the planet. Metastable helium atoms are colored as a function of their radial velocity in the stellar rest frame (see color bar). The dashed circle is the projected transition between the exosphere and thermosphere regimes. All particles in this projected view are outside of the thermosphere. The eccentric orbit of HAT-P-11b (black disk, plotted above the exosphere for the sake of clarity) is shown as a green curve. The tail is due to particles in the planet shadow being protected from photo-ionization. *Panels b and c*: View along the LOS (line of sight) toward Earth, showing the thermospheric (b) and the exospheric (c) regimes separately. The thermosphere (b) is colored as a function of the column density of metastable helium. Particles in the exosphere (c) are colored as in panel (a). Panel b shows the grids discretizing the stellar disk, the thermosphere, and the planetary disk.

**Acknowledgments:** We acknowledge the Geneva exoplanet atmosphere group for fruitful discussions and the support of X. Dumusque for the DACE platform.

**Funding:** This work was based on observations collected at the Centro Astronómico Hispano Alemán (CAHA), operated jointly by the Max-Planck Institut für Astronomie and the Instituto de Astrofísica de Andalucia (CSIC) under OPTICON program 2017B/026, 'Multi-wavelength observations of the warm Neptune HAT-P-11b: a journey across the atmosphere'. This project has received funding from the European Union's Horizon 2020 research and innovation programme under grant agreement No 730890. This material reflects only the authors views and the Commission is not liable for any use that may be made of the information contained therein. This work has been carried out in the frame of the National Centre for Competence in Research 'PlanetS' supported by the Swiss National Science Foundation (SNSF). R.A., V.B., D.E., C.L., L.P., F.P. and A.W. acknowledge the financial support of the SNSF. A.W. acknowledges the financial support of the SNSF by the grant number P2GEP2_178191. A.L.E thanks the CNES for financial support. This project has received funding from the European Research Council (ERC) under the European Union's Horizon 2020 research and innovation programme (project FOUR ACES; grant agreement No 724427). This project has also received funding from the European Research Council (ERC) under the European Union's Seventh Framework Programme (Fp7/2007-2013) / ERC grant agreement no. 337592.

**Author contributions:** R.A. coordinated the study, performed the data reduction and analyzed the results. V.B. developed the EVE code, based on previous code written by V.B. and A.L.E.. R.A. and V.B. performed the EVE simulations and wrote the paper. J.J.S. performed the *HST* and *JWST* simulations. R.A., V.B., C.L., D.E., L.P., A.W. and F.P. wrote the OPTICON proposal. All authors participated in the discussion and interpretation of the results, and commented on the manuscript.

**Competing interests:** The authors declared no competing interests.

**Data and materials availability:** The CARMENES raw and reduced data can be obtained from the Calar Alto Observatory archive at http://caha.sdc.cab.inta-csic.es/calto/jsp/searchform.jsp under programme number 051. The data and simulation outputs used to produce each figure are available at doi:10.5281/zenodo.1473463 . The EVE code is described in (*13*, *18*, *28*).

**Supplementary Materials**

Materials and Methods

Table S1 – S2

Fig S1 – S6

References (30 - 44)

# Supplementary Materials for
## Spectrally resolved helium absorption from the extended atmosphere of a warm Neptune-mass exoplanet

R. Allart[1]*, V. Bourrier[1], C. Lovis[1], D. Ehrenreich[1], J.J. Spake[2], A. Wyttenbach[1,3], L. Pino[1,4,5], F. Pepe[1], D.K. Sing[2,6], A. Lecavelier des Etangs[7]
Correspondence to: romain.allart@unige.ch

**This PDF file includes:**

Materials and Methods
Figs. S1 to S6
Tables S1 to S2

**Materials and Methods**

PHYSICAL PARAMETERS

Orbital and physical parameters of the HAT-P-11b system have been determined by radial velocity and transit techniques (*1–3*). The detection of the secondary eclipse with the *Kepler* satellite (*3*) brought strong constraints on parameters useful to our study (orbital period $P$, mid-point of the transit $T_0$, eccentricity $e$ or argument of the periastron $\omega$, Table S1). In addition we refined the mass of HAT-P-11b, using the DACE (Data & Analysis Center for Exoplanets) platform (*30, 31*) to analyse the existing radial velocity data obtained with the HIRES spectrograph (High Resolution Echelle Spectrometer) mounted on the Keck telescope (*1*). We applied a Metropolis–Hasting Markov chain Monte Carlo algorithm with Gaussian priors based on the results from (*3*). Our best-fitting values for the stellar semi-amplitude $K_*$, the stellar mass $M_*$, the planetary mass $M_\mathrm{p}$, the semi-major axis $a$, $e$ and $\omega$ are shown in Table S1.

DATA REDUCTION

We used the data reduction applied automatically by the CARACAL (CARMENES Reduction And Calibration, (*10*)) pipeline. It consists in a bias, flat and cosmic ray correction of the raw spectra, followed by a flat-relative optimal extraction (FOX; e.g., (*11*)) and wavelengths calibration (*12*). The resulting output is in the Earth laboratory frame and is composed of wavelength (in vacuum), flux and flux uncertainty maps (order vs. pixel number).

We excluded from our analysis spectra with a signal to noise ratio below 50 per pixel (of which there were 7 in Visit 1, 8 in Visit 2) while the median signal to noise ratio is 87 and 96 per pixel for Visit 1 and 2. These spectra show narrow emission features of unknown origin at ~10,835 Å that are seen only at high airmass. About 30 minutes were lost in visit 2 due to technical problems. We also excluded from Visit 2 the spectrum taken at 20:48 UT (Universal Time), because car lights induced spurious emission at ~6,600 Å and ~12,800 Å. The dataset used in our analysis thus consists of 53 and 51 spectra, each of 408 s duration, representing a total of 6 h and 5.8 h of observations. 19 exposures in Visit 1 and 18 in Visit 2 were obtained during the planetary transit (duration 2.4h), with a baseline, respectively 29 exposures (before transit) and 20 (before and after transit) to perform transmission spectroscopy.

We focused our analysis on the helium triplet, using the spectra derived from order 55 between 10,804 Å and 11,005 Å. Between 10,827 Å and 10,840 Å (Fig. S1), near the helium triplet, we identified telluric OH emission lines at 10,834.4, 10,834.7 and 10,836.6 Å using a comparison sky fibre, a strong telluric water line at 10,837.4 Å overlapping with the latter OH line, two weak water lines at 10,827.4 Å, 10,839.3 Å, and a bad pixel at 10831 Å. We used Molecfit (*32*) to correct for telluric water lines in absorption following the same procedure as (*16*). Most absorption telluric lines are corrected to the noise level. The strong water telluric line at 10,839.3 Å is not well corrected due to contamination by the nearby OH emission line, which is not taken into account in Molecfit. OH emission lines have not been corrected as they are at least 1.8 Å away from the planetary signature (between 10,831.5 and 10,832.7 Å in the star frame) and thus do not contaminate our detection. The bad pixel at 10,831 Å was corrected by fitting a fourth-order polynomial to neighboring pixels. None of these telluric and detector features directly overlap with the helium triplet. During Visit 1, an emission feature is observed in the core of the deep stellar line at ~10,829.9 Å. It becomes more prominent towards the end of the night. This feature does not correspond to known water or OH transitions. Fig. S2 shows that it arises from the observer's or star frame rather than the planetary frame because the line is narrower and stronger in this frame than in the planetary frame. To avoid biasing the numerical fits applied to the transmission spectra (see below), we replaced the 4 affected pixels with the mean value of the corresponding pixels in Visit 2.

IMPACT OF ORBITAL ECCENTRICITY

We used the following Keplerian radial velocity model to correct the stellar and planetary reflex motions:

$$V_{\text{rad}} = K \cdot [\cos(\theta + \omega) + e \cdot \cos(\omega)] \quad \text{Eq. S1}$$

For a circular orbit, the radial velocity of the planet during a transit will span a symmetrical range centered at 0 km·s$^{-1}$ (e.g. HD189733b's radial velocities range from –15 km·s$^{-1}$ to 15 km·s$^{-1}$, (*14*)). HAT-P-11b has an eccentricity of 0.26 and its radial velocity during the transit increases from –36 to –24 km·s$^{-1}$ (–30 km·s$^{-1}$ at mid-transit). In the case of an eccentric orbit, the stellar and planetary absorption lines can be easily distinguished in velocity space (Fig. S2). The eccentric orbit of HAT-P-11b blueshifts its atmospheric absorption signature into a region of the stellar spectrum with no strong absorption lines. The projected rotational velocity of HAT-P-11 is 1 km·s$^{-1}$, and the orbit of the planet is nearly polar with an impact parameter of 0.35 (*33*), which implies that it occults stellar surface regions with similar low radial velocities of about 200 m·s$^{-1}$. Therefore, we do not expect important contamination from the Rossiter-McLaughlin effect, and it was not taken into account in our analysis. The atmospheric signal is strongly diluted if studied in the stellar frame. During the transit the planet radial velocity changes by tens of kilometers per second. If one co-adds all the individual transmission spectra in the stellar frame, the planetary absorption feature will be strongly diluted in wavelength resulting in a weaker and broader line compared to the planetary frame (Fig. S2).

TRANSMISSION SPECTRUM AND LIGHT CURVE

CARMENES spectra were shifted to the stellar rest frame by correcting them for the barycentric Earth radial velocity (BERV), the systemic velocity of the star and its reflex motion induced by the planet. Each spectrum taken at phase below -0.0125 (-1.5 h) or above 0.024 (+2.8 h) is considered an out-of-transit spectrum, to avoid contamination by the planetary atmosphere or exosphere (e.g. cometary like-tail). Then we co-added all the out-of-transit spectra ($\lambda$ is the wavelength and *t* is the time from the mid-point of the transit), $f(\lambda, t_{out})$, to build the normalized master out-of-transit spectrum, $F_{out}(\lambda) = \sum f(\lambda, t_{out})$. All individual spectra were normalized to the continuum level of the master out-of-transit spectrum with a fourth-degree polynomial (*16*). We applied a sigma-clipping rejection algorithm on the spectra in order to replace all the cosmic ray hits by the median value of the other spectra at the same wavelength. Then, we computed the integrated transmission spectrum, $\Re(\lambda)$, as the sum of each individual transmission spectrum, $f(\lambda, t_{in})/F_{out}(\lambda)$, after applying a Doppler shift to the planet rest frame $p$ to compensate for the orbital motion during the transit.

$$\mathfrak{R}(\lambda) = \sum_{t \in in} \left. \frac{f(\lambda, t_{in})}{F_{out}(\lambda)} \right|_p \quad \text{Eq. S2}$$

We finally expressed the variation of the occulted area by the planet as a function of wavelength on an absolute scale (*14–16*), by using the reference white-light radius ratio, $\frac{R_p^2(\lambda_{ref})}{R_*^2}$ (Table S1):

$$\frac{R_p^2(\lambda)}{R_*^2} = 1 - \mathfrak{R}(\lambda) + \frac{R_p^2(\lambda_{ref})}{R_*^2} \quad \text{Eq. S3}$$

We also computed the helium absorption light curve. Ground-based spectra can only measure the excess absorption because absolute fluxes are lost due to absorption in Earth's atmosphere. We derived quadratic limb-darkening coefficients from the transit photometry of HAT-P-11b in J band (from 1.05 to 1.34 µm) to compute the theoretical white-light curve (*34*). We used it to scale the individual transmission spectra. Then, fluxes are integrated from 10,832.84 Å – 10,833.59 Å in each individual transmission spectrum and compared to reference band passes at 10,825 Å – 10,828.3 Å, and 10,856.7 Å – 10,860.2 Å to obtain the total helium absorption light curve from HAT-P-11b and allow comparison with simulations of its atmosphere (see below).

THE EVAPORATING EXOPLANET CODE

The EVE code and its application to hydrogen and helium atmospheres has been published and described in (*18, 28, 35*). It calculates realistic theoretical spectra as they would be measured with a given instrument during the transit of an exoplanet. The direct comparison of these theoretical spectra with observations allows us to constrain planetary and stellar properties. The code has been applied to a variety of systems (e.g. (*18*, *35*)): and we describe here the general workings of its latest version. The settings and inputs of the EVE code relevant to its application to HAT-P-11b are detailed in the next section.

A planetary system is simulated in 3D in the star rest frame, the planet moving along a predetermined orbit and its atmosphere described using two different regimes. Monte-Carlo particle simulations are used to compute the dynamics of atoms and ions escaping into the exosphere, while a parameterized grid is used for the atmospheric layers between the exosphere and the planet surface (or a reference pressure level). Observations fitted with the EVE code usually trace the hot, extended upper atmospheric layers immediately below the exosphere, hence we consider that the parameterized grid describes the thermosphere of the planet. At each time step of the simulation, the code computes the absorption of the intrinsic stellar spectrum by the planetary disk, the thermosphere, and the exosphere. The stellar disk is discretized so that the code processes the absorption of the local spectrum emitted by each stellar cell. At the end of the simulation, the time-series of disk-integrated spectra are resampled over the spectral and temporal dimensions to match observed spectra, and then processed with the relevant instrumental response. We describe these different steps in more details below.

**Thermosphere regime**

The thermosphere, found at pressures lower than 1 mbar, is the part of the atmosphere where extreme UV radiation (FUV (Far-UltraViolet), XUV), X-rays and cosmic rays are absorbed, and part of their energy is converted to heat. As a result, temperature increases with altitude in the lower thermosphere, leading the upper thermospheric layers to lose their stability and expand hydrodynamically. Temperature decreases as the gas expands in the upper atmosphere, ultimately fueling escape of atomic species at the exobase (at pressures lower than 1 nbar) (*36*). The simulated thermosphere is described with a 3D cubic-cell grid, defined with respect to the stellar grid in order to facilitate the calculation of the theoretical spectra (see below). The 2D cells of the stellar grid are discretized with a square grid at the resolution chosen for the thermospheric cells. Each of these sub-cells can be seen as the base of a column parallel to the line-of-sight (LOS) between the Earth and the star, which is further discretized along this direction. At each timestep, the code identifies the cells in each column that are contained within the thermosphere spherical boundaries. It then retrieves the properties of the thermospheric cells along each column to calculate their spectral opacity profile. The simulated thermosphere can include multiple species, defined by their density, velocity, and temperature, all of which can vary along the three spatial dimensions. The structure of the thermosphere is parameterized, i.e. it can be defined using either analytical functions or tabulated profiles (defined on the basis of the studied planet), rather than calculated self-consistently.

**Exosphere regime**

The transition between the thermospheric and exospheric regimes (the thermopause, or exobase) is intended to represent the altitude up to which the atmosphere keeps a global cohesion. Particles from different species can be released at the thermopause, and their dynamics is then calculated into the exospheric regime. Individual atoms are too numerous to be modeled individually, and they are thus grouped into metaparticles representing groups of atoms with the same properties. The number of atoms in a metaparticle is defined so that one metaparticle yields a low elementary opacity in front of a stellar disk cell and within a spectral bin of the stellar spectrum. The number of metaparticles launched at each timestep is controlled by an escape rate parameter. Metaparticles can be released locally or over the entire thermopause, and their launch velocity relatively to the planet is set by a chosen distribution, typically a thermal component from a Maxwell-Boltzman velocity distribution combined with a constant radial bulk component.

The code calculates the dynamics of all species metaparticles in the stellar reference frame, accounting for the planet and star gravity, the inertial force linked to the non-Galilean stellar reference frame, and the relevant physical mechanisms (e.g. the stellar radiation pressure or interactions with the stellar wind). Metaparticles can further be subjected to changes of state, in particular photoionization. Particles in the shadow of the planetary disk are not subjected to any processes arising from stellar particles (photons or wind particles). When required the code can process self-shielding of photons and stellar wind particles by the planet atmosphere, which reduces their impact on exospheric layers farther away from the star. Particles ending a timestep within the thermosphere boundary or within the star because of their dynamics are removed from the simulation.

**Calculation of theoretical spectra**

A reference spectrum is used to define the local intrinsic spectrum emitted by each cell of the stellar grid, taking into account the effective area of the star surface in each cell, and the local limb-darkening. The disk-integrated spectrum of the star is then recalculated as the sum of the local intrinsic spectra from all cells. At each simulation timestep, the code calculates the attenuation of the local stellar spectrum in front of each cell, accounting for the opaque planetary disk occultation and for the wavelength-dependent absorption from the thermosphere and the exosphere regimes. The disk-integrated spectrum received from the planetary system at a given timestep thus writes:

$$F(\lambda) = \sum_{Star\ cells} F(\lambda, r) \cdot \left(1 - \frac{S_p(r)}{S_*(r)} - \sum_{\substack{Columns \\ in\ cell}} \frac{S_c}{S_*(r)} \cdot \left(1 - e^{-\Sigma_{subcells}\ \tau_c(\lambda)}\right)\right)$$

$$\cdot\ e^{-\Sigma_{Particles\ in\ cell}\ \tau_p(\lambda)} \qquad \text{Eq. S4}$$

where $F(\lambda, r)$ is the local spectrum from a stellar cell at the sky-projected distance r from the star center. $S_*(r)$ is the effective area of the stellar disk within a cell of the stellar grid. $S_p(r)$ and $S_c$ are the occulting surfaces of the planetary disk and of a thermospheric column in front of a given stellar cell. Thermospheric columns and metaparticles in front of or behind the opaque planetary disk do not contribute to the absorption. The planet is discretized with a square grid of 700 km wide cells, so that $S_p(r)$ is calculated by summing the number of occulting planet-grid cells.

$\tau_c(\lambda)$ and $\tau_p(\lambda)$ are the spectral opacity profiles from a cell in a thermospheric column and from a metaparticle, respectively. The cross-section profiles depend on the species, the absorption process (electronic transition, broadband absorption, etc.), and the atmospheric regime. Voigt profiles are typically used for the cross-section of gas in the thermosphere, with thermal broadening and spectral position determined by the local cell temperature and radial velocity. In the exosphere regime, the cross-section of metaparticles associated to atomic species are described with Lorentzian profiles centered on the particle's radial velocity.

When processing a given stellar cell, the code loops over all occulting thermospheric columns to calculate their cumulated contribution to the absorption profile of the stellar cell spectrum (Eq. S4). For each column, the code sums the opacity profiles of all relevant absorption processes, after interpolating them over the wavelengths of the reference stellar spectrum (opacity profiles are calculated over a spectral table with resolution equal or larger than the resolution of the reference stellar spectrum). Once planetary occultation and thermospheric absorption have been processed for a given stellar cell, the code calculates the global opacity profile from all metaparticles occulting the cell (Lorentzian cross-section profiles are integrated analytically, directly over the bins of the reference stellar spectrum table).

Eq. S4 yields the raw theoretical spectrum received from the planetary system at given timestep. At the end of the simulation, raw spectra are convolved with the chosen instrumental response and interpolated over a spectral table adjusted so that each observed spectral bin contains an integer number of adjusted raw bins. The adjusted raw spectra are then averaged within each observed spectral bin. The resulting spectra are further averaged within the time window of the observed exposures, weighting their relative contribution by the effective overlap with the observed time window. These final theoretical spectra are comparable to observed spectra: they are 'exposed' over the same time windows, defined over the same wavelength table, and they account for the response of the instrument. Before they can be compared, the observable and observed spectra might however need further processing to account for the effects of Earth atmosphere.

APPLICATION OF THE EVE CODE TO HAT-P-11B

We now describe the specific application of the EVE code to the case of helium in the upper atmosphere of HAT-P-11b. There are three features in the simulations of HAT-P-11b that require a realistic stellar spectrum at high-energy and in the region of the helium triplet: the calculation of theoretical spectra to be compared with the CARMENES observations, the calculation of stellar radiation pressure on metastable helium atoms, and the calculation of their photo-ionization rates. The MUSCLES (Measurement of the Ultraviolet Spectral Characteristics of Low-mass Exoplanetary Systems) database (*37*) provides a library of observed and synthetic stellar spectra from the X-ray to the infrared. We retrieved the MUSCLES spectra ('adapt-var-res-sed.fits' products, version 2.1) of HD 85512 (K6-type) et HD 40307 (K2.5-type) to build a proxy for HAT-P-11 (K4-type). The two spectra were rescaled in flux at 1 astronomical unit (au) from their respective stars, and averaged. While the resulting spectrum was found to be in good agreement with the black body spectrum of HAT-P-11 (stellar radius $R_* = 0.68\,R_\odot$, effective temperature $T_{eff} = 4{,}780$ K), it did not provide a good match to the unocculted stellar spectrum measured with CARMENES in the region of the He I triplet. We thus used the CARMENES spectrum in this region to get a more accurate calculation of radiation pressure. The final combined spectrum was oversampled by a factor of 6 (pixel size = 0.01 Å, equivalent to 0.27 km·s$^{-1}$ at 10,833 Å), with respect to the CARMENES spectra, and used to define the local intrinsic stellar grid spectra in the simulations. Limb-darkening over the stellar grid was calculated using coefficients defined in the region of the helium triplet (Table S1). Spectral opacity profiles from the thermosphere and the exosphere were calculated at a resolution of 2 km·s$^{-1}$, and we checked that decreasing this value did not change our results. All calculations were done using wavelengths in vacuum.

**Thermosphere regime**

The planet below the thermosphere was simulated using the orbital and stellar properties in Table S1. We found that the overall shape of HAT-P-11b helium absorption signature could be well reproduced by an isotropic, hydrostatic density profile in the thermosphere. The actual profile is likely more complex, but this simple structure is a good approximation for the expanding atmospheres of giant planets (*38*) and allows us to derive first-order properties of HAT-P-11b's upper atmosphere using a minimal number of free parameters. There is a degeneracy between $T_{th}$ and $\mu$, and we thus defined the profile using the parameter $T_{th}/\mu$ rather than assuming a specific composition. We included in the thermospheric grid a constant upward velocity $v_{th}$ (ie, along the normal to the planet surface) to represent the bulk expansion of the upper thermosphere driven by the high-energy stellar XUV radiation absorbed in the lower thermosphere. The projection of the upward velocity upon the LOS can contribute to the broadening of the atmospheric absorption signature.

**Exosphere regime**

We assume that the different species escaping from the thermosphere of HAT-P-11b are decoupled, and that they can be modeled independently. Once helium atoms escape into the exosphere the local densities are likely too low (by definition the exosphere is a collision-less regime) for new metastable helium atoms to be formed via recombination of excited helium atoms with electrons, or via collisions of ground-state helium atoms with electrons (*17*). Therefore, we only model the population of escaping metastable helium atoms. The number of atoms in a metaparticle was set to $1 \cdot 10^{27}$, yielding elementary opacity on the order of $1 \cdot 10^{-3}$ for the strongest transition. Because the orbit of HAT-P-11b is eccentric, the mass loss rate of metastable helium was set to vary as the inverse square of the distance to the star, with the model value $\dot{M}_{\mathrm{He}^*}$ defined at the semi-major axis. At each timestep, the code scales the density profile of metastable helium in the thermosphere so that it matches the density of metaparticles measured in a shell just above the exobase, set at a radius $R_{\mathrm{trans}}$.

HAT-P-11 naturally exerts a radiation pressure on metastable atoms in the exosphere, which can be described as a force opposite stellar gravity (*39*). The ratio $\beta$ between the two forces is proportional to the stellar flux received by an atom at the wavelengths of its absorption lines. Radiation pressure is thus wavelength-dependent and calculated by the code using the disk-integrated stellar spectrum. The total radiation pressure coefficient on a given metaparticle is the sum of the coefficient $\beta$ from the three helium lines, calculated at the radial velocity of the particle with respect to each line center (Fig. S3). A K-type star like HAT-P-11 has a bright continuum in the near-infrared, and the helium triplet lines have high oscillator strengths (up to 0.3). This results in a cumulative radiation pressure reaching up to about 90 times the stellar gravity (Fig. S3). The stellar flux, and hence the radiation pressure, is roughly constant blueward of the helium transitions, resulting in a constant acceleration on escaping particles. A deep stellar absorption line stops this acceleration once particles reach about −90 km·s$^{-1}$.

Metastable helium atoms in the simulation can be photo-ionized by the stellar incident radiation, or radiatively de-excite into their ground state. Both processes subject the helium population to exponential decay with rates $\alpha_{ion}$ and $\alpha_{deexc}$, respectively. Photoionization rates are calculated for each metaparticle using the wavelength-dependent cross-section for metastable helium (*40*), and the disk-integrated stellar spectrum up to the photoionization threshold (2,593 Å). The de-excitation rate corresponds to a constant decay lifetime of 7,870 s (*41*). At each time step $dt$ of the simulation, we randomly choose whether the chance that a given metaparticle is photoionized or de-excites, using the probability $P = 1 - \exp(-(\alpha_{ion} + \alpha_{deexc}) \cdot dt)$, and if so the particle is removed from the simulation. We do not account for self-shielding of the stellar photons by the exosphere, because its dominant constituent is expected to be hydrogen whose neutral state would absorb the stellar light in the Lyman-alpha line alone. Simulations were performed with a timestep of 30 s to resolve the dynamics and photoionization of the metastable.

**Comparison of EVE and CARMENES spectra**

Theoretical spectra are processed as described above to allow comparison with the CARMENES spectra, assuming the instrumental response is a Gaussian with Full width at Half Maximum, *FWHM* = 3.7 km·s$^{-1}$. Earth's atmosphere induces a global variation in the flux measured with CARMENES during a night, leading (a) to variations in the distribution of flux with wavelength that can be different for each exposure, and (b) to the loss of absolute flux levels. Effect (a) is compensated for in the observations by calculating transmission spectra and normalizing them to a constant continuum level, as described in the main text. We calculated equivalent theoretical transmission spectra by normalizing EVE spectra with the disk-integrated spectrum convolved with CARMENES instrumental response. These theoretical transmission spectra show the continuum absorption from the simulated planetary disk, which is lost in the observed spectra because of effect (b). We thus calibrated the observed transmission spectra using HAT-P-11b continuum transit light curve (see main text).

We used the $\chi^2$ as merit function to compare the theoretical and observed transmission spectra, generating a grid of simulations as a function of the four free parameters in the model: $T_{th}/\mu$, $v_{th}$, $R_{trans}$ and $\dot{M}_{He^*}$. We explored a domain of the parameter space wide enough that it encompasses all models within at least 3-sigma from the best-fitting values (Table S2). We limited our exploration to exobase altitudes within the Roche Lobe radius (6.5 $R_p$), as the atmosphere would likely deviate from a spherical shape at larger altitudes. The altitude of the exobase could be derived from the knowledge of the total gas density in the thermosphere, however the CARMENES observations are only sensitive to the density of metastable helium and there are no observational constraints on the density of other species in the upper atmosphere of HAT-P-11b. Hydrodynamical simulations of a large sample of planets' upper atmospheres yield maximum temperatures of about 16,000 K, and neutral hydrogen fractions down to a few percent for the hottest thermospheres (20). With a pure hydrogen composition, this would correspond to a mean atomic weight of about 0.5, which we can take as an absolute lower limit on $\mu$. We thus limited our exploration to $T_{th}/\mu \leq 16,000 / 0.5$ = 32,000 K·amu$^{-1}$.

Once the grid of models was explored, we analyzed the variations of the $\chi^2$ as a function of each parameter independently to identify the range of values within 1-sigma from the best-fitting model (Fig. S4, Table S2). The comparison of the best-fitting absorption profile with the observed signature revealed that it is symmetrical but significantly blueshifted. While a blown-away helium exosphere could potentially yield an apparent blueshift of the total absorption profile (because of its excess absorption at negative velocities), there are not enough helium atoms escaping from the model thermosphere to produce such an effect. 2D hydrodynamical simulations of extended atmospheres (*21*) predict the formation of winds flowing from the day- to the night-side up to high altitudes, yielding a net blueshift of the atmospheric absorption profile. We thus included in our atmospheric model of HAT-P-11b a constant zonal wind and performed simulations over a range of wind velocities while fixing the other model parameters to their best-fitting values (Fig S4). Zonal wind speeds of ~3 km·s$^{-1}$ were found to reproduce the blueshift of the observed absorption profile, while contributing negligibly to its shape dominated by thermal broadening. The contribution of zonal winds is thus not degenerate with that of the other model parameters. This justifies a posteriori our assumption that the influence of zonal winds could be investigated independently from the other model parameters.

COMPARISON WITH *HST* AND *JWST*

We compared our CARMENES detection to simulated observations with the Wide Field Camera 3 (WFC3) on board of *HST* and the Near Infrared Spectrograph (NIRSpec) on board of *JWST*. We used a model stellar spectrum from the Phoenix grid (*42*) which most closely matches the stellar parameters for HAT-P-11 ((*1*); Table S1). To model the planetary transmission spectrum, we added the best-fitting EVE model of the helium absorption to a model transmission spectrum for HAT-P-11b from a grid of ATMO models (*43*) degraded to a resolution of 200 and 2700, respectively for *HST/* WFC3 and *JWST*/NIRSpec. To estimate the expected uncertainty on the absorption signal for HAT-P-11b with *HST*, we re-scaled the error bars from a measured transmission spectrum of WASP-107b using WFC3 with the grism G102 (*28*). We accounted for the difference in host star brightness (J = 9.4 for WASP-107, and 7.6 for HAT-P-11), and assumed uncertainties of 1.17 times the photon noise for 98 Å - wide channels, as was measured with WASP-107 with one transit observation (*28*). With an optimal exposure time of 60 s and an out-of-transit baseline equal to the transit duration (2.3 hours), we predict a detection with a significance of 2.4σ.

We used the PandExo noise simulator (*44*) to estimate the expected uncertainties on the 10,833 Å excess absorption with *JWST*/NIRSpec using the grism G140H and the filter F070LP mode, which gives the highest spectral resolution around 10,833 Å ($\Delta\lambda < 3$ Å). We computed for a single transit, with an out-of-transit baseline equal to the transit duration (2.3 hours), exposure times of 0.45 s with 3 groups per integration. For this setup the predicted uncertainty on a single-pixel channel is 110 ppm at 10,833 Å, which leads to an expected significant detection of 60 σ in the 1-pixel channel centered on the line.

Therfore, *HST*/WFC3 and *JWST*/NIRSpec cannot resolve the helium triplet, but *JWST*/NIRSpec has a higher sensitivity and higher resolution than *HST*/WFC3 that allow measuring the atmospheric helium triplet at higher confidence. High-resolution spectrographs provide complementary information into the atmospheric thermal and dynamical structure by spectrally resolving the helium lines.

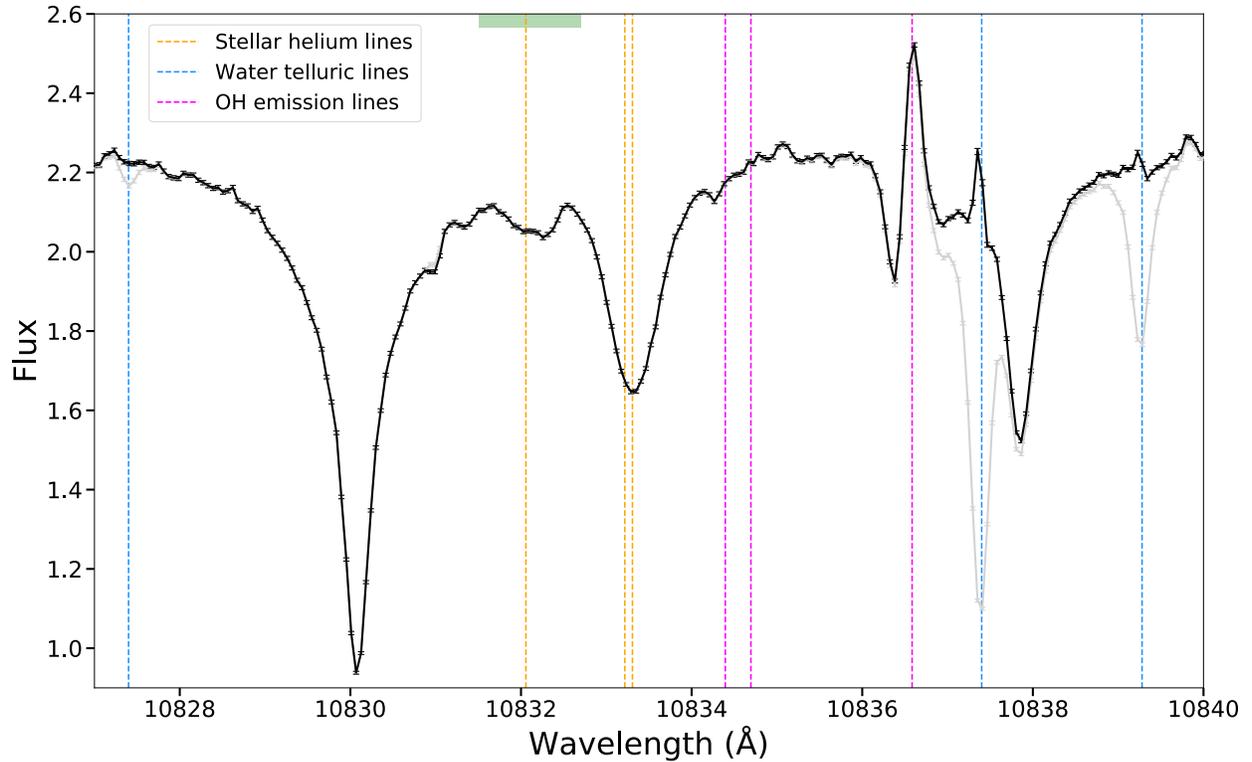

**Fig. S1. Master out-of-transit spectrum in the stellar rest frame for visit 2.** In grey, the master spectrum before telluric correction and in black after telluric correction by Molecfit. The telluric water lines (dashed blue vertical lines) at 10,827.1 Å and 10,839.3 Å are well corrected, while the water line at 10,837.9 Å shows residuals, most likely due to the presence of an OH emission line at 10,836.6 Å. Dashed magenta vertical lines correspond to OH lines observed in the observer rest frame as seen in simultaneous observations of the blank sky. Dashed orange vertical lines correspond to the rest wavelengths of the He I triplet. The planetary absorption helium lines span from 10,831.5 Å to 10,832.7 Å in the stellar rest frame (green band).

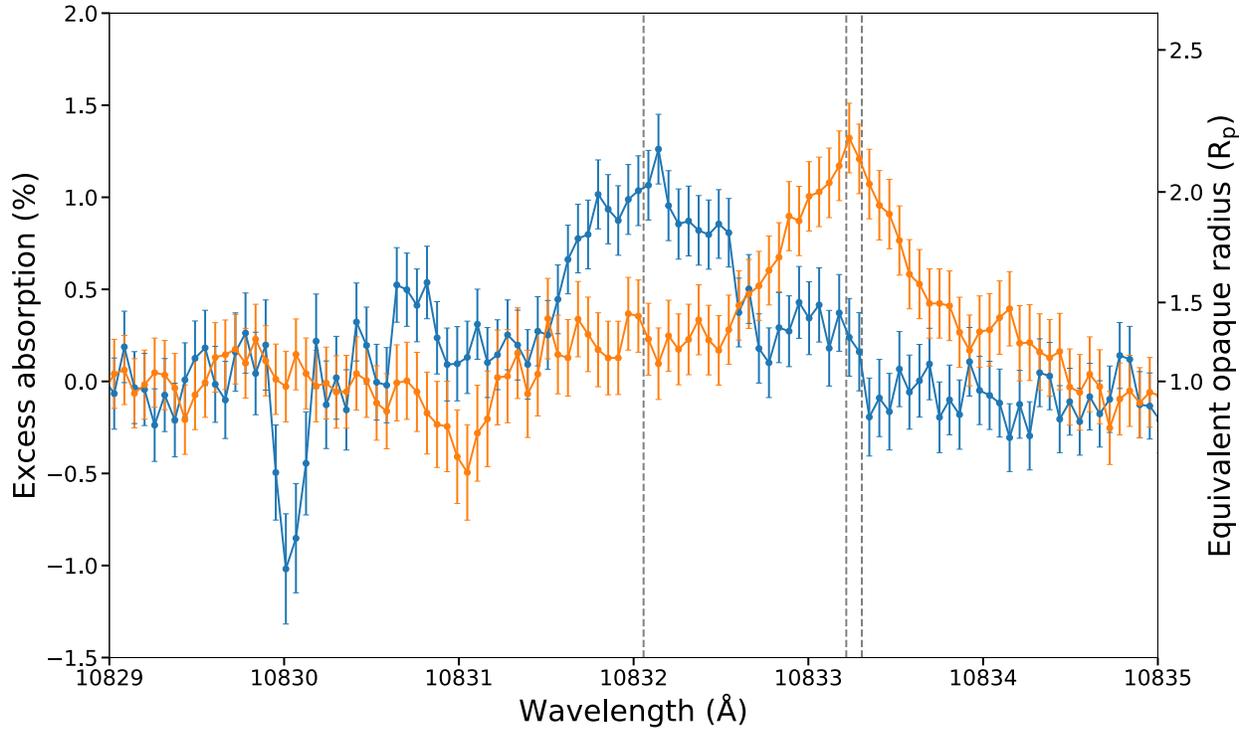

**Fig. S2. Transmission spectrum in the stellar (blue) and planetary frame (orange).** The unknown emission feature at 10,830 Å is narrower and stronger in the stellar frame, which is linked to the observer's frame by a constant offset. The eccentricity of HAT-P-11b allows us to differentiate between stellar and planetary frames: the transmission spectrum in the planet frame shows helium absorption that is well centered on the rest wavelengths (dashed grey vertical lines), while the transmission spectrum in the stellar frame is shifted relative to the rest wavelengths. The absorption in the star frame appears diluted and slightly weaker compared to the planetary frame.

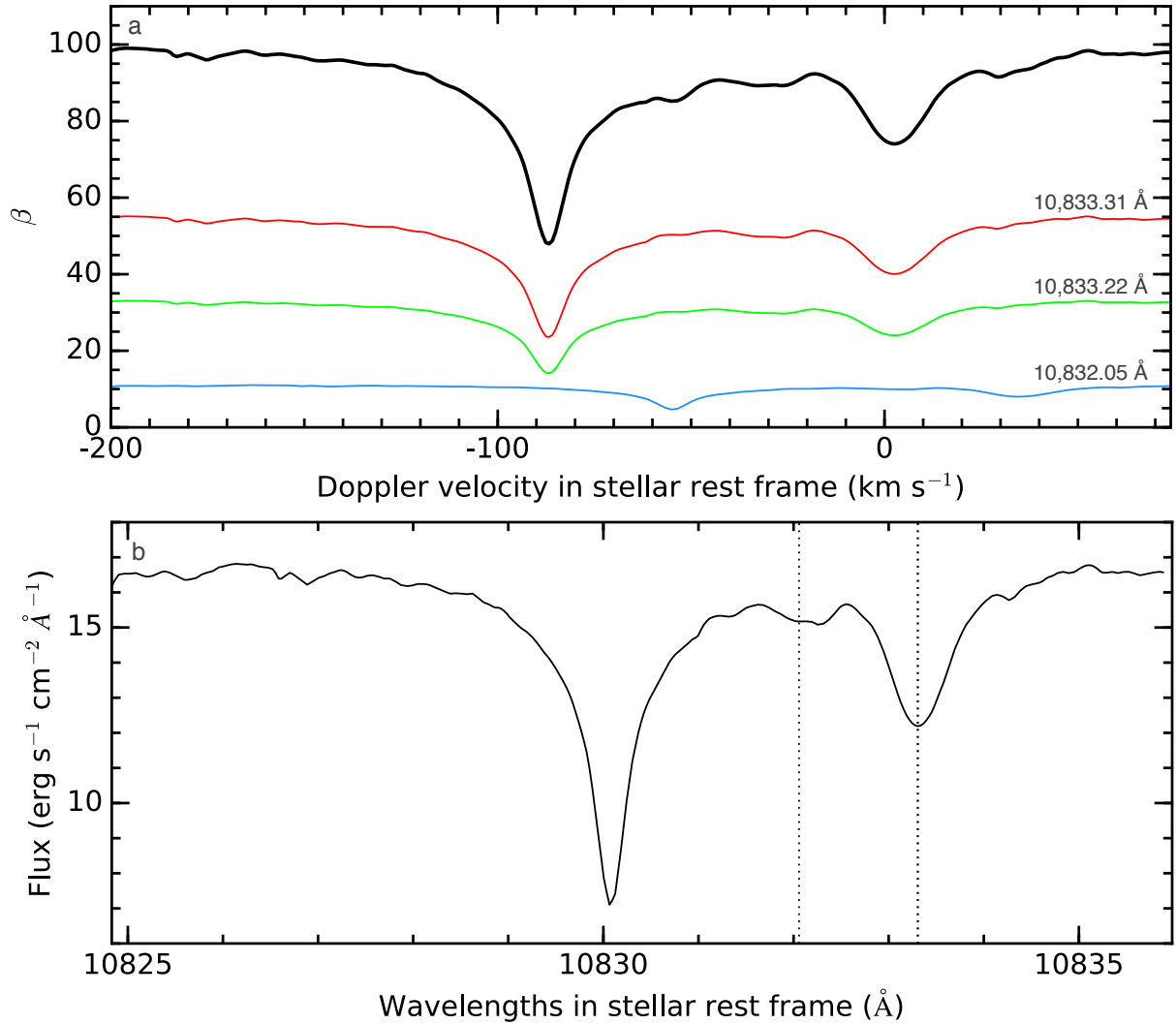

**Fig. S3. Radiation pressure from the He I triplet.** *Panel a:* Radiation pressure coefficients associated with the He 10,832.05 Å line (blue line), the He 10,833.22 Å line (green line), and the He 10833.31 Å line (red line), as a function of radial velocity in the star rest frame. Their sum $\beta$ (solid black line) characterizes the total radiation pressure profile seen by a metastable helium atom. *Panel b:* Intrinsic stellar flux at 1 au from HAT-P-11, in the spectral range where it contributes to the radiation pressure profiles shown in the upper panel. Vertical dashed lines show the helium triplet line transitions.

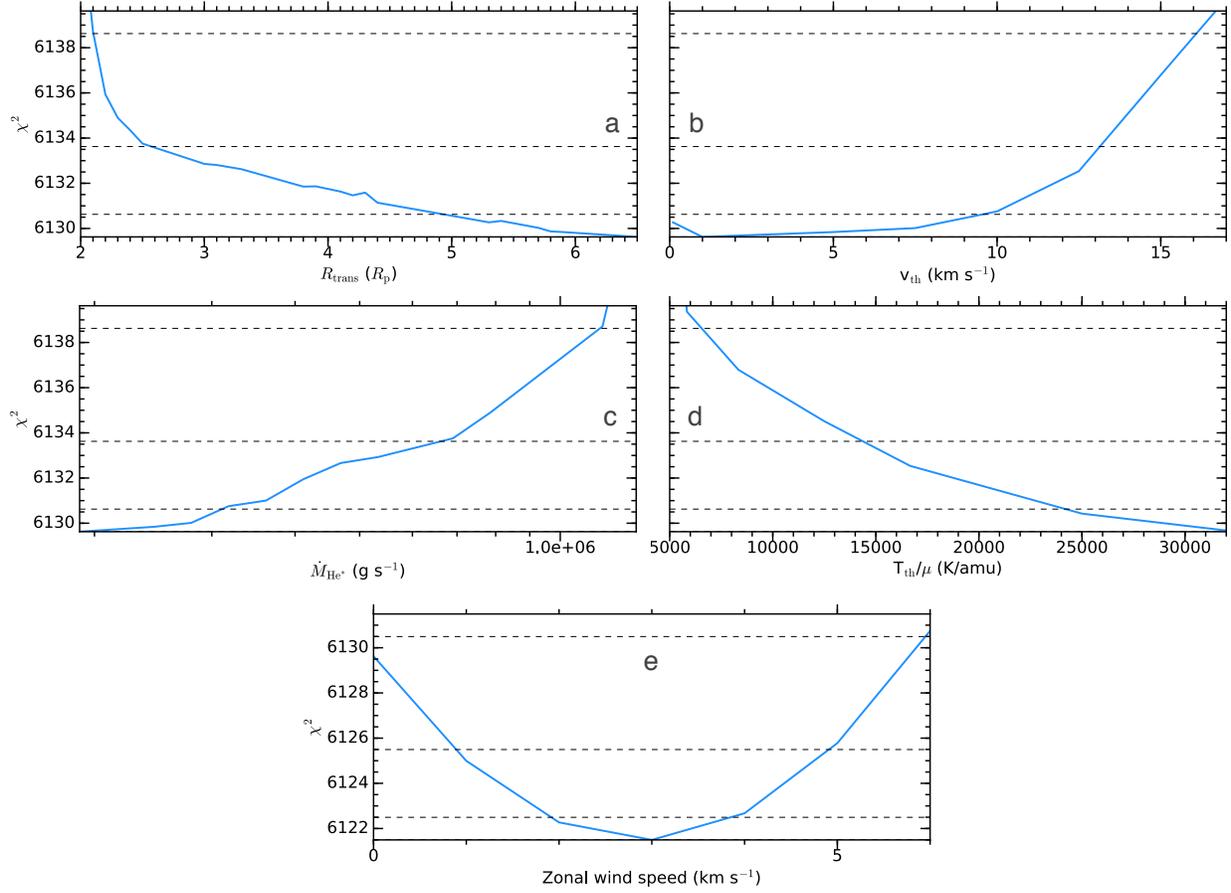

**Fig. S4. $\chi^2$ variations for each free parameter in the model.** Models in the panel a, b, c and d do not include the contribution of zonal winds. The panel e shows variations of the $\chi^2$ as a function of zonal wind speed, with the other model parameters fixed to their best-fitting values. Dotted horizontal lines indicate $\chi^2$ values at 1, 2, and 3 σ from the best-fitting.

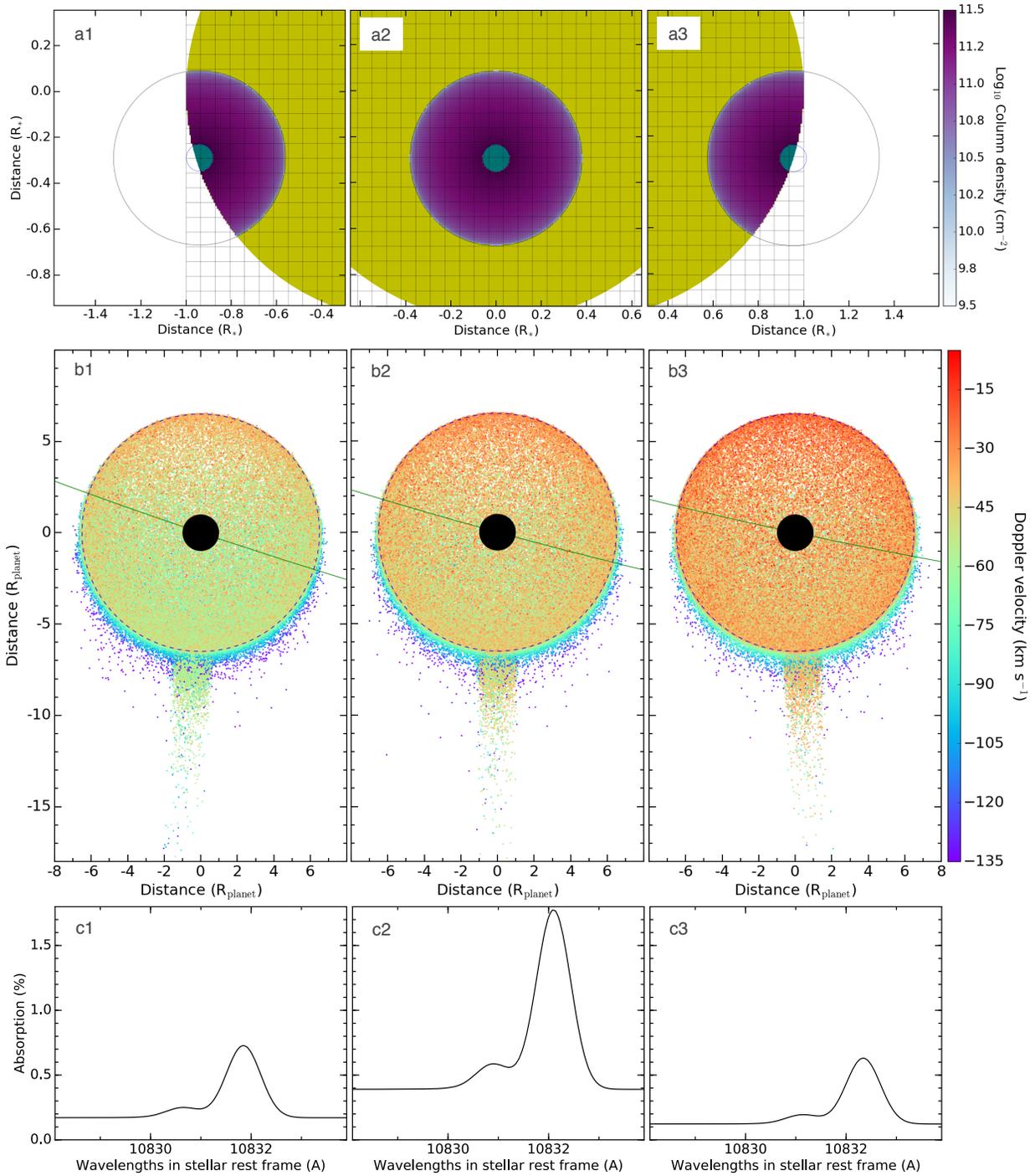

**Fig. S5. EVE best-fitting simulation of HAT-P-11b transit in the He I triplet.** The three columns show the system simulated during ingress (t=-1h7min, panels 1), at mid-transit (t=0h, panels 2), and during egress (t=1h8min, panels 3). *Panels a:* Column density grid associated to the thermosphere regime, similar to Fig. 4b. *Panels b:* View from the above of the planetary system,

similar to Fig. 4a, showing metaparticles in the exosphere regime colored as a function of their radial velocity along the LOS. The star is toward the top of the image, and the plot is centered on the planet (plotted as a black disk). *Panels c:* Theoretical raw absorption profiles as a function of wavelength in the star rest frame, illustrating the spectral shift induced by the eccentric orbital motion of HAT-P-11b.

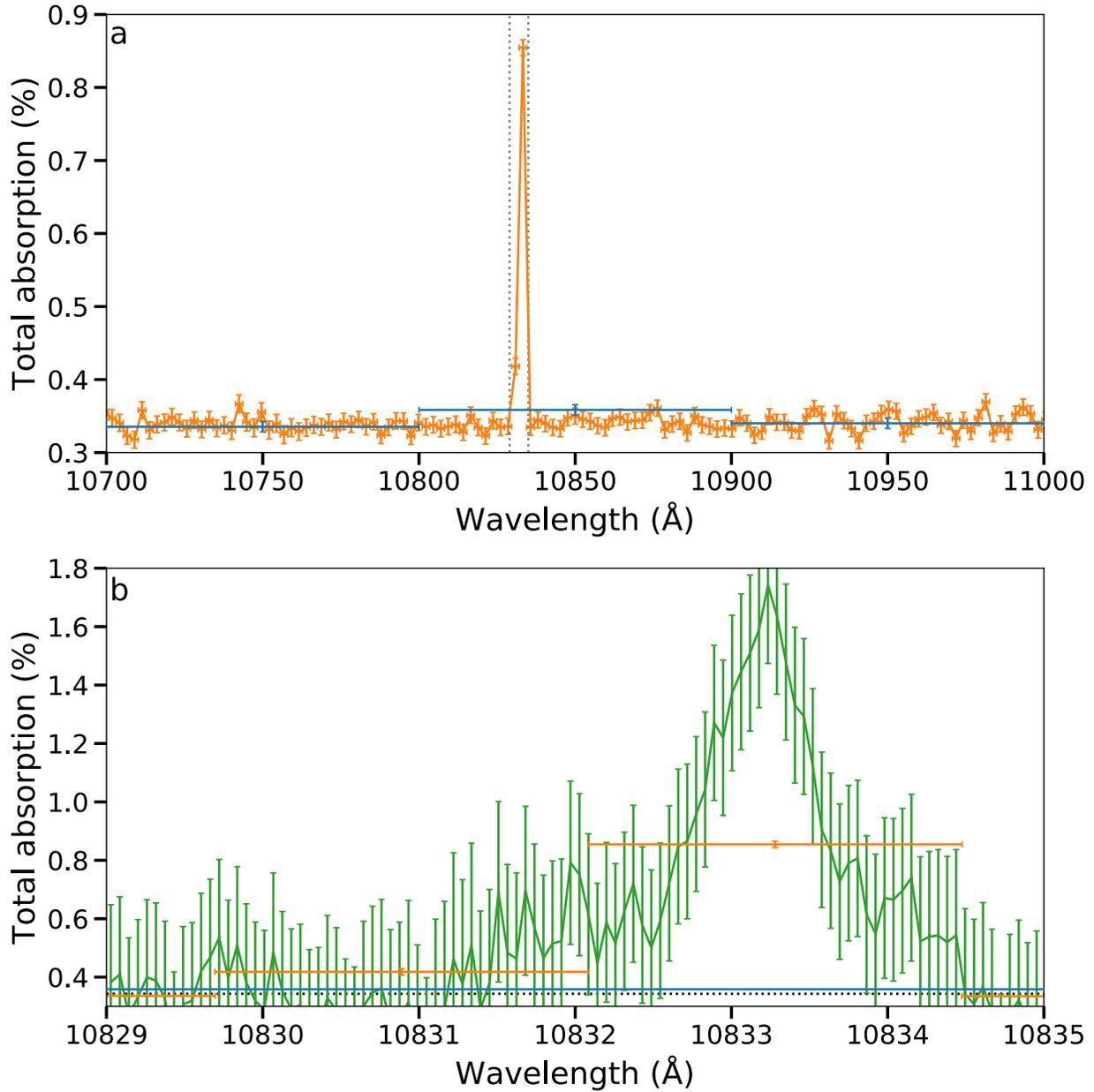

**Fig. S6. Comparison between the CARMENES (green), and simulated *JWST*/NIRSpec (orange) and *HST*/WFC3 (blue) transmission spectra.** *Panel a:* Simulated transmission spectra with the space-based instruments *JWST*/NIRSpec and *HST*/WFC3, in the range of the helium triplet. Grey vertical dotted lines correspond to the spectral range of panel b. *Panel b:* Close-up view on the resolved absorption signature measured with CARMENES and comparison with the other instruments. The black dotted line corresponds to the opaque disk. The He I triplet is resolved

with CARMENES while it is covered by one- and two- pixels channel of *HST*/WFC3 and *JWST*/NIRSpec, respectively.

**Table S1. Physical and orbital parameters of the HAT-P-11 system.** $M_\odot$ is the Solar mass. $R_\odot$ is the Solar radius. $M_\oplus$ is the Earth mass. $R_\oplus$ is the Earth radius.

| Quantity | Symbol | Values | References |
|---|---|---|---|
| Stellar mass | $M_*$ | $0.802 \pm 0.028\ M_\odot$ | This work (DACE) |
| Stellar radius | $R_*$ | $0.683 \pm 0.009\ R_\odot$ | (2) |
| Planetary mass | $M_p$ | $27.74 \pm 3.11\ M_\oplus$ | This work (DACE) |
| Planetary radius | $R_p$ | $4.36 \pm 0.06\ R_\oplus$ | (3) |
| White light radius ratio | $R_p(\lambda_{optical})/R_*$ | $0.05850 \pm 0.00013$ | (3) |
| Orbital period | $P$ | $4.887802443 \pm 0.000000034$ days | (3) |
| Mid-point of the transit | $T_0$ | $2454957.8132067 \pm 0.0000053$ days (BJD) | (3) |
| Transit duration | $t_{14}$ | $0.0981875 \pm 0.0000625$ days | (3) |
| Semi-amplitude | $K_*$ | $12.01 \pm 1.38\ \mathrm{m\cdot s^{-1}}$ | This work (DACE) |
| Systemic velocity | $\gamma$ | $-63.428 \pm 0.003\ \mathrm{km\cdot s^{-1}}$ | This work (DACE) |
| Semi-major axis | $a$ | $16.50 \pm 0.18\ R_*$ | This work (DACE) |
| Orbital inclination | $i$ | $89.05 \pm 0.15\ °$ | (3) |
| eccentricity | $e$ | $0.264353 \pm 0.000602$ | This work (DACE) |
| Argument of the periastron | $\omega$ | $342.185794 \pm 0.179084\ °$ | This work (DACE) |
| Temperature effective | $T_{\mathrm{eff}}$ | $4780 \pm 50$ K | (1) |
| Metallicity | [Fe/H] | $0.31 \pm 0.05$ | (1) |
| Surface gravity | $\log(g)$ | $4.59 \pm 0.03$ (cgs) | (1) |
| distance | $d$ | $38 \pm 1.3$ pc | (1) |
| Linear limb darkening | $u_1$ | 0.267 | Exofast (34) |
| Quadratic limb darkening | $u_2$ | 0.265 | Exofast (34) |

**Table S2. Physical and orbital parameters of the HAT-P-11 system explored in the simulation.** Limits on $T_{th}/\mu$ were imposed based on physical justification. The other boundaries were not defined a priori, but define the parameter space explored within 3-sigma from the best-fitting model

| Quantity | Symbol | Explored range |
| --- | --- | --- |
| Thermosphere scale height parameter | $T_{th}/\mu$ | 6,000 – 32,000 K·amu$^{-1}$ |
| Thermosphere upward velocity | $v_{th}$ | 1 - 20 km·s$^{-1}$ |
| Radius of the exobase | $R_{trans}$ | 1.5 – 6.5 R$_p$ |
| Escape rate of metastable helium | $\dot{M}_{He^*}$ | $10^4 – 10^7$ g·s$^{-1}$ |